\documentclass[aps,prb,twocolumn]{revtex4}
\usepackage{pstricks,epsfig,subfigure,changebar,picinpar,rotating,graphicx}
\def\ad{{a_{\vec k}^\dag}}
\def\a{{a_{\vec k}}}

\def\ap{{a_{\vec k'}}}
\def\bd{{b_{\vec k}^\dag}}
\def\b{{b_{\vec k}}}
\def\bdp{{b_{\vec {k'}}^\dag}}
\def\bp{{b_{\vec{k'}}}}

\def\omkp{{\omega (\vec k')}}
\begin{document}

\date{\today }
\title{Control of non-Markovian effects in the dynamics of polaritons in semiconductor microcavities}
\author{F.J.Rodr\'{\i}guez and L.Quiroga}
\affiliation{Departamento de F\'{\i}sica, Universidad de los
Andes, A.A. 4976, Bogot\'a D.C., Colombia}
\author{C.Tejedor}
\affiliation{Departamento de F\'{\i}sica Te\'orica de la Materia
Condensada, Universidad Aut\'onoma de Madrid, Cantoblanco,
E-28049, Madrid, Spain}
\author{M.D.Mart\'{\i}n and L. Vi\~na}
\affiliation{Departamento de F\'{\i}sica de Materiales C-IV,
Universidad Aut\'onoma de Madrid, Cantoblanco, E-28049, Madrid,
Spain}
\author{R.Andr\'e}
\affiliation{Lab.Spectrom\'etrie Physique (CNRS), Universit\'e
Joseph Fourier 1, F-38402 Grenoble, France}
\begin{abstract}
We report on time-resolved photoluminescence from semiconductor
microcavities showing that an optically controllable mechanism
exists to turn on and off memory effects in a polariton system. By
increasing the laser pumping pulse intensity we observe revivals of
the decaying time-resolved photoluminescence signal, a manifestly
non-Markovian behavior of the optically active polaritons. Based on
an open quantum system approach we perform a comprehensive
analytical and numerical study of the coupling of optically active
polaritons to a structured reservoir to confirm the origin of the
observed features. Our findings show that negative detunings and
strong excitation should occur simultaneously for memory effects to
take place.

\end{abstract}
\maketitle
\section{Introduction}
Semiconductor microcavities (SM) have attracted a great deal of
attention in recent years due to the opportunity they bring to
create and manipulate, in a controlled way, many-bosons systems in a
solid-state-environment\cite{savvidis,baumberg}. Bose-Einstein
condensation signatures of radiation-matter quasiparticles
(polaritons) have been recently identified in such
systems\cite{dang1,dang2,balili,lai}. On the other hand, due to the
rich possibilities of tailoring the matter-radiation interaction in
SM, optically controlled dynamics of elementary electronic
excitations is within reach. In particular, the understanding of the
ultrafast dynamics of polaritons becomes crucial for interpreting
important quantum control experiments in SM as it has been
demonstrated recently\cite{savasta}. A convenient and versatile way
of monitoring the ultrafast dynamics of polaritons is provided by
time-resolved photoluminescence (tr-PL) experiments.

Here we focus our attention on II-VI SM where a strong
exciton-LO-phonon coupling produces a rapid relaxation for
non-resonantly created polaritons, with large excess
energies\cite{alexandrou}. It has been reported that, under certain
conditions, tr-PL following a non-resonant pulsed excitation of a
CdTe-based microcavity shows an oscillatory emission dynamics
strongly depending on the detuning and the initial excitation
density. Furthermore, spin-related effects have been observed when
the tr-PL is analyzed into its co- and cross-polarized components
after excitation with circularly polarized
pulses\cite{vina,lola,vinaand}. The non-linear coupling of optically
active and dark states has been invoked as a possible mechanism to
explain existing experimental results\cite{shelykh}. However, these
unusual experimental features seem still to challenge conventional
theoretical approaches.

Most theoretical models rely on the Born-Markov approximation to
describe the polariton dynamics\cite{ciuti,porras1,porras2}. The
main feature of these approaches is to neglect memory effects, that
is, the behavior of the polariton system at some time $t$ is only
determined by its configuration precisely at the same time. The
validity of this assumption requires the environment characteristic
correlation time to be small as compared with the relaxation time of
the system. The fingerprint of a Markov process is an exponential
decay, while deviations from a Markovian behavior cause a
non-exponential time evolution, with eventually superimposed
oscillations implying some degree of correlation between the quantum
system and its environment.

The main purpose of the present paper is to describe an optically
controllable mechanism of memory effects in SM and to report on
tr-PL results which provide its quantitative verification. Several
parameters may be used to control the presence or absence of memory
effects in the dynamics of a polariton system. Some of them are of a
static nature, such as the detuning between the cavity optical mode
and the bare exciton resonance, whereas others are of a dynamic
nature, such as the intensity of the optical pumping. Here we
emphasize on the importance of the latter ones.

The analysis is carried out within the framework of the theory of
open quantum systems. In the usual scenario for studying open
quantum systems, a central system of interest is {\it directly}
coupled to a large system usually labeled as the bath. In contrast,
here we consider instead a three systems framework: the quantum
system of interest (the emitting polaritons), a high-energy
(exciton-like) polariton bath and an intermediate system (at the
"bottleneck" region). The last system provides control over the {\it
indirect} coupling between optically active polaritons and the bath
of high energy polaritons. We show that, initial excitation
intensity behaves as a controllable memory mechanism for producing a
rich variety of features in tr-PL signals. Spin effects are not
included in the present study.

The paper is organized as follows: in Section II we present our
three coupled systems model. We start by briefly discussing some
aspects of a master-like equation approach with memory effects that
may provide insights on the proper dynamics of the emitting
polariton system. However, polariton-polariton interactions together
with polariton-loss effects are hard to include and thus make this
simple approach unpractical to provide an appropriate description of
the problem. In order to obtain a good quantitative agreement with
experimental data a microscopic Heisenberg-Langevin approach,
including memory effects, has been adopted \cite{viena}. In Section
III we present and discuss the experimental data, which corroborate
our theoretical predictions. A summary is presented in Section IV.

\section{Theoretical Background}
\subsection{The model}
After their creation in the upper polariton (UP) branch by a pulsed
laser, polaritons rapidly relax to the lower polariton (LP) branch
states. Once in the LP branch, one of the main mechanisms for
polariton scattering in SM is the extremely efficient
parametric-down-conversion. By this process two polaritons are
simultaneously scattered: one of them goes to a
$\overrightarrow{k}\sim 0$ or "signal" particle state while the
other one (conserving energy and linear momentum) goes to a high
$\mid \overrightarrow{k}\mid$ exciton-like state, the so-called
"idler" particle state. Thus, we start by identifying three regions
of interest in the LP branch, as sketched in Figure 1: (i) The
bottom of the trap (signal states), $\overrightarrow{k}\sim 0$, with
energy dispersion $E(\vec k)$, described by operators $\a$; (ii) the
intermediate or "bottleneck" polariton region, described by
operators $c_{\vec k}$ and energy dispersion $\epsilon(\vec k)$,
where polaritons accumulate after a rapid relaxation from the UP
branch, and finally (iii) the exciton-like bath (idler states)
described by operators $b_{\vec k}$ and energy dispersion
$\omega(\vec k)$. The Hamiltonian is then given by ($\hbar=1$)
\cite{porras1}
\begin{eqnarray}
\nonumber H&=&\sum_{\vec k} E(\vec k)\ad\a + \sum_{\vec k} \omega(\vec k) \bd
\b + \sum_{\vec k} \epsilon(\vec k) c_{\vec k}^{\dag}c_{\vec k}\\
&+&H_{XP}+H_{SI}\label{Eq:e1}
\end{eqnarray}
In the present model polaritons are to be described by boson
operators linearly interacting with a large boson bath (excitons).
The polariton-polariton interaction term is given by
\begin{eqnarray}
H_{XP}=\sum_{\vec k,{\vec k}^{\prime},{\vec k}_1,{\vec
k}_2}\tilde{V}(\vec k,{\vec k}^{\prime},{\vec k}_1,{\vec k}_2)\ad
b_{{\vec k}^{\prime}}^{\dag}c_{{\vec k}_1}c_{{\vec k}_2}+H.C.
\end{eqnarray}
where $\tilde{V}(\vec k,{\vec k}^{\prime},{\vec k}_1,{\vec k}_2)$
accounts for the polariton-polariton coupling and $H.C.$ means
hermitic conjugate. Since a large population of polaritons can
condense at the bottom of the trap in the LP branch,
$\overrightarrow{k}\sim 0$, a self-interaction term has to be
included as given by
\begin{eqnarray}
H_{SI}=V_0a_0^{\dag}a_0^{\dag}a_0a_0
\end{eqnarray}

\begin{figure}[tbh]
\centerline{ {\includegraphics [height=10.5cm,width=9.0cm]{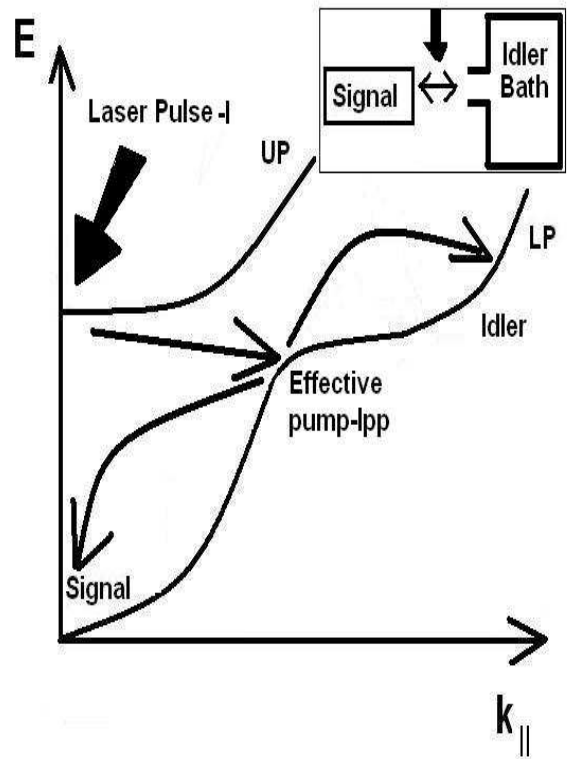}}
} \caption{\label{fig-1} Schematic representation of different
polariton subsystems. Inset: two systems, signal and idler,
indirectly coupled by an intensity dependent strength.}
\end{figure}

Intermediate quasiparticles ("bottleneck" polaritons), can reach a
high population due to rapid relaxation processes from high energy
states. Thus, the effective interaction between optically active
polaritons and exciton-like polaritons in the bath can be assumed to
be modulated by the presence of those "bottleneck" quasiparticles.
It is well known that for composite environments, where extra
degrees of freedom modulate the interaction between a quantum system
of interest and a large reservoir, an effective non-Markovian
behaviour on the quantum system dynamics arises even when the
reservoir itself can be described in the Markovian approximation
\cite{budini}. Following this line of thought, we shall consider
that polaritons optically pumped at the UP branch rapidly accumulate
into an intermediate or "bottleneck" region of
quasiparticles\cite{bloch}.

The exact quantum evolution of the fast relaxing polariton system is
far from describable in simple terms and a detailed theory of its
evolution is not yet available. We can, however, fill this gap with
the reasonable hypothesis that, under high intensity pumping, the
intermediate polaritons can be described by a parametric
approximation by using classical fields instead of full quantum
field operators. This approximation ignores quantum fluctuations in
the intermediate polariton fields. Thus, in $H_{XP}$, the $c_{{\vec
k}_1}c_{{\vec k}_2}$ operator is replaced by the c-number $
<c_{{\vec k}_1}c_{{\vec k}_2}>\sim h_{{\vec k}_1}(t)\delta_{{\vec
k}_1,{\vec k}_2}$ with $h_{{\vec k}}(t)$ acting as an effective pump
on the intermediate polaritons, depending on the actual pumping
pulse at the UP branch. In this context, an effective classical
intensity sets the system-reservoir coupling strength. Consequently,
the polariton-polariton interaction term adopts the time-dependent
effective form
\begin{eqnarray}
H_{XP}(t)=\sum_{\vec k,\vec k'} V(\vec k,\vec k',t) \ad\bdp
+\sum_{\vec k,\vec k'} V^*(\vec k,\vec k',t) \a\bp\label{Eq:e4}
\end{eqnarray}
where
\begin{eqnarray}
V(\vec k,\vec k',t)=\sum_{{\vec k}_1,{\vec k}_2}\tilde{V}(\vec
k,\vec k',{\vec k}_1,{\vec k}_2)h_{{\vec k}_1}(t)\delta_{{\vec
k}_1,{\vec k}_2}
\end{eqnarray}
Pump depletion is well accounted for with a simple-minded pulse
shape as determined by the shape/length of the excitation laser
pulse as well as by proper polariton relaxation mechanisms in a
microcavity. We shall return to this point later. The effective
Hamiltonian describing the coupling of trapped polaritons ($a_{\vec
k}$ modes) with the high energy polariton bath ($b_{\vec k}$ modes),
Eq.(\ref{Eq:e4}), corresponds to a coupling strength $V(\vec k,\vec
k',t)$, which is now adjustable experimentally by varying the
excitation pulse parameters such as their width and intensity. Thus,
it is already clear from the present discussion how an optically
controlled mechanism may exist to turn on and off memory effects in
a polariton system.

Additionally, the above description explains why a Born-Markov based
theoretical approach should be invalid in the present context: (i)
even if the optically active polariton system is weakly coupled to
the high energy polariton reservoir, a {\it time-dependent} coupling
strength yields to a non-exponential decay; (ii) by continuously
increasing the pump pulse intensity a strongly coupled
system-reservoir situation should be reached beyond a certain
threshold, hence a theory based on a lowest-order perturbation in
the coupling strength breaks down; (iii) the idler-polariton bath,
as being formed by higher energy polaritons (massive exciton-like
particles) with a quadratic dependence of $\omega(k)$ on $k$,
constitutes a highly structured continuum, for which a single
characteristic correlation time cannot be identified.

It is worth noting that the Hamiltonian in Eq.(\ref{Eq:e1}), with
$H_{XP}$ as given by Eq.(\ref{Eq:e4}), corresponds to a
non-degenerate parametric amplifier for massive particles, where the
parametric gain implies that the LP population, as  well as the
polariton bath population, are amplified at the expenses of
intermediate polariton depletion. We emphasize that we are
considering a pulsed excitation, thus no indefinite gain will arise
and the validity of the parametric approximation is still
guaranteed. Furthermore, for a time-independent system-reservoir
coupling, our problem can be solved exactly. We use this as a check
on our numerical results later.

Before we address the full numerical solution of the proposed model,
an analytically tractable method is to be discussed first. This
discussion is presented here because it complements the results of
the microscopic treatment to be detailed below and it will also
serve as a reference point.

\subsection{Time-convolutionless non-Markovian master equation}
The first type of approximation we consider is based in a
semi-analytical approach by using a Lindblad-like master equation
with time-dependent rates allowing non-Markovian effects to be
included. We start by considering a bipartite system conformed on
one hand by the optically active polaritons $\overrightarrow{k}\sim
0$ and on the other hand by a high energy polariton bath. The
underlying physics is analogous to the parametric-down conversion
processes in non-linear optics, hence consequently with an {\it
anti-rotating-wave}-like coupling between those sub-systems. For an
initially empty polariton bath, the master-like equation for the
reduced density operator of optically active (signal) polaritons,
without any other decay mechanism, is
\begin{eqnarray}
\nonumber &\frac{\partial \rho_S(t)}{\partial t}&=-\frac{i}{2}S(t)[a_0^{\dag}a_0,\rho_S(t)]+\\
&\frac{1}{2}\gamma(t)&\left ( -a_0a_0^{\dag}\rho_S(t)-
\rho_S(t)a_0a_0^{\dag} +2a_0^{\dag}\rho_S(t)a_0 \right )
\label{Eq:etcl}
\end{eqnarray}
Time-convolutionless master equations of the latter form have been
largely documented\cite{libro}. The time-dependent parameters $S(t)$
and $\gamma(t)$ depend on the reservoir spectral density
\cite{breuer}. Eq.(\ref{Eq:etcl}) is local in time but contains all
the information about memory effects in the time dependent parameter
$\gamma(t)$.

From this master equation is immediate to obtain the equation of
motion for $n(t)=Tr_S\{ \rho_S(t) a_0^{\dag}a_0 \}$, the
population of optically active polaritons
\begin{eqnarray}
\frac{d n(t)}{d t}=\gamma(t)(n(t)+1)
\end{eqnarray}
Given that initially the ${\vec k}=0$ mode is empty, $n(0)=0$, we
found that the number of optically active polaritons grows up as
\begin{eqnarray}
n(t)=e^{\int_{0}^{t}\gamma(t')dt'}-1
\end{eqnarray}
In the Markov regime for which $\gamma(t)=\gamma_M=$constant, the
polariton population grows exponentially. However, non-Markovian
effects are evident when the relaxation rates in the master equation
are time dependent. Up to now, no dissipation effects have been
included in our discussion. Thus, the polariton population in the
optically active mode does not cease to steadily increase.
Polariton-polariton interactions, creating scattered quasiparticles
out of the $\overrightarrow{k}\sim 0$ zone, cavity losses and
recalling that the system-bath coupling is pulsed for a finite
period of time (no continuous pumping mechanism replenishing the
intermediate polariton zone is present), produce finally that $n(t)$
temporally saturates and then goes to 0. Nevertheless, any
oscillation associated with the time-dependent rate $\gamma(t)$
should manifest itself not only in the rise- but also in the
decay-evolution of the optically active polariton density.

One noteworthy feature of this treatment is the possibility of
clarifying the growth of $n(t)$ at early times. In the Markov
approximation, $n(t)=e^{\gamma_Mt}-1$, thus at short times the
polariton population starts growing with a linear slope. Since a
signature of non-Markovian behavior is a time dependent relaxation
rate, which starts growing as $\gamma(t)\sim gt$, the polariton
population should behave at short times as $n(t) \sim
e^{gt^2/2}-1\sim gt^2/2$, in qualitative agreement with the
experimental results referring to the curvature of the initial time
evolution (see below).

Despite the exact solvability of this model a systematic approach
for dealing with mechanisms of polariton interactions and losses are
not easily implementable, which makes the present formalism
inadequate to describe the non-monotonic dependence of tr-PL
signals. We consider, therefore, an analytical solution of the
microscopic Heisenberg dynamics to investigate in a more detailed
way memory effects in polariton systems.

\subsection{Heisenberg-Langevin dynamics}
From the previous analysis, it is apparent that there is a large
amount of theoretical insight to be gained from a more exhaustive
examination of the proposed model. Now, a thorough formal analysis
of the proposed polariton dynamics is performed. We solve directly
the Heisenberg equation of motion for a given operator $A$. In
particular, for $\a$ and $b_{\vec k}^{\dag}$ we obtain
\begin{eqnarray}
i {\dot a}_{\vec k}&=&E(\vec k) \a + \sum_{\vec k'} V(\vec k,\vec
k',t) b_{\vec k'}^\dag + 2 V_o a_0^\dag a_0 a_0 \delta_{\vec
k,0}\label{Eq:e6}
\\
i {\dot b}_{\vec k}^{\dag}&=&-\omega(\vec k)\bd-\sum_{\vec k'}
V^*(\vec k,\vec k',t) \ap \label{Eq:e7}
\end{eqnarray}

In order to solve numerically the coupled equations of motion,
Eqs.(\ref{Eq:e6}) and (\ref{Eq:e7}), we formally integrate $\bd(t)$
in Eq.(\ref{Eq:e7}), then inserted in Eq.(\ref{Eq:e6}), to get
\begin{eqnarray}
\nonumber &\dot \a&(t)=-i E(\vec k)\a - i\sum_{\vec k'} V(\vec
k,\vec
k',t)\bp(0)e^{i\omkp t}\\
&+&\nonumber \int_0^t\sum_{\vec k',\vec k''}V(\vec k,\vec k',t)
V^*(\vec k',\vec k'',t-\tau)a_{\vec k''}(t-\tau)e^{i\omkp
\tau}d\tau \nonumber
\\
&-&2 i V_0 a_0^\dag a_0 a_0 \delta_{\vec k,0} \label{integro}
\end{eqnarray}
For the sake of simplicity we take $\vec k'' =\vec k$ and assume
that the effective interaction $V$ can be separated as $V(\vec
k,\vec k',t)=g(\vec k,\vec k') h(t)$, where $g(\vec k,\vec k')$
accounts for both the Coulomb and Pauli effects in the
polariton-polariton scattering while $h(t)$, a dimensionless
function of time, represents an effective pump-polariton pulse. As a
consequence, Eq. (~\ref{integro}) can be rewritten as
\begin{eqnarray}
\dot \a(t)&=&-i \left( E(\vec k)-i\Gamma_0 +2V_0 a_0^{\dag} a_0
\delta_{\vec k,0}\right)\a-i\eta_{\vec k}(t) \nonumber
\\
&+& \int_0^t h(t)h(t-\tau)\a(t-\tau)K_{\vec
k}(\tau)d\tau\label{Eq:ef}
\end{eqnarray}
in terms of the kernel function
\begin{eqnarray}
K_{\vec k}(\tau)=\sum_{\vec k'}g(\vec k,\vec k') g(\vec k',\vec k)
e^{i\omega(\vec k')\tau} \label{Eq:ker}
\end{eqnarray}
and a polariton-bath noise function
\begin{eqnarray}
\eta_{\vec k}(\tau)= \sum_{\vec k'}g(\vec k,\vec k')\bp(0)
e^{i\omega(\vec k')\tau}h(\tau)
\end{eqnarray}
Since the memory time of the radiation field outside the microcavity
is extremely short, on the order of $1/E(0)\sim 1$ fs for typical
II-VI gap energies $E(0)\sim 2-3$ eV, we are justified to treat
radiation losses from the microcavity within the Markov
approximation with the simple inclusion of a phenomenological
damping term $\Gamma_0$ in Eq.(\ref{Eq:ef}). However, of primary
interest here are the non-Markovian effects coming from the strong
coupling between signal and idler-bath polaritons. Eq.(\ref{Eq:ef})
embodies the memory effects on the dynamics of the emitting
polaritons. The term $g(\vec k,\vec k')$ describes the strength and
spectral form of the signal-idler coupling.

We emphasize that, in our model, the bath is formed by high energy
polaritons (exciton-like quaiparticles), hence the idler-bath we are
considering consists of massive particles. In this sense, the
present signal-idler polariton coupled system is very similar to
atom-laser systems with a continuous output
coupler\cite{breuer,lambropoulos}. However, the main difference with
atom-lasers is that in our case the signal-idler outcoupling is of
an anti-rotating-wave kind (see Eq.(\ref{Eq:e4})) instead of the
standard rotating-wave approximation usually employed in atomic
systems. In order to proceed, we assume a quadratic energy
dispersion relation for the idler particles, $\omega(k)=k^2/2M_X$
with $M_X$ the bare exciton effective mass, and a Gaussian-like
profile for the trapped ground-state polariton in $k$-space.
Consequently, the signal-idler coupling $g(\vec k,\vec k')$ can be
written (in the continuum approximation for idler particles) as
\cite{breuer,lambropoulos}
\begin{eqnarray}
g(\vec k,\vec
k')=\frac{i\Gamma^{1/2}}{(2\pi\sigma_k^2)^{1/4}}e^{-\frac{(\vec k -
\vec k')^2}{4\sigma_k^2}}
\end{eqnarray}
where $\Gamma^{1/2}$ and $\sigma_k$ settle the strength and width of
the system-reservoir coupling, respectively. The kernel term, for
the optically active mode $k\sim 0$,  becomes
\begin{eqnarray}
K_{0}(\tau)=\frac{\Gamma}{\sqrt{1-i\alpha\tau}} \label{Eq:kernel}
\end{eqnarray}
where $\alpha=\frac{\sigma_{0}^2}{M_x}$. Note that in our model, the
bath is assumed to be formed by bosons with a dispersion relation
$\omega(k)=k^2/2M_X$. This fact gives rise to distinct features in
the spectral bath response and consequently in the reduced system
dynamics, as compared with that found for non-massive and
structureless baths, such as those corresponding to photons or
phonons.

The noise term is the responsible for initiating the polariton
relaxation towards the bottom LP states. Thus, we adopt for this
term the role of a classical seed and have checked numerically
that a very small value for it does not affect the results.

At this point we have completed the presentation of our theoretical
framework and the underlying approximations, thus we can now proceed
to evaluate the time evolution of the mean number of emitting
polaritons at the bottom of the lower polariton branch to compare it
with tr-PL experimental data.

\section{Results}

\subsection{Experiments}

The sample under study is a Cd$_{0.4}$Mg$_{0.6}$Te $\lambda$-cavity,
with top (bottom) distributed Bragg reflectors (DBRs) built with
17.5 (23) pairs of alternating $\lambda$/4 thick layers of
Cd$_{0.4}$Mg$_{0.6}$Te and Cd$_{0.75}$Mn$_{0.25}$Te. In each of the
antinodes of the electromagnetic field confined in-between the DBRs
there are two CdTe quantum wells (QWs) of 90{\AA} thickness. The
strong coupling between the excitons confined in the QWs and the
photons confined in the cavity yields to a Rabi splitting of
$\sim$10 meV at low temperature. One important parameter to be
considered is the detuning $\delta=E(0)-\omega(0)$. The cavity
thickness varies across the wafer, allowing to tune the photon in
and out of resonance with the excitons, thus varying the detuning.

The sample is kept inside a cold-finger cryostat at a temperature of
$~8$ K and is resonantly excited with 2 ps-long pulses arriving at
the sample at $\sim 3^o$, their energy tuned to the UP branch. The
time evolution of the PL is obtained by means of a spectrograph
coupled to a streak camera (time resolution $\sim$10 ps). We have
selected the emission originating from $k\sim0$ lower polariton
states by means of a small pinhole (angular resolution $\sim 1^o$).
For polarization-resolved measurements we have used two
$\lambda$/4-plates to excite and analyze the PL into its
co/cross-circularly polarized components, after excitation with
$\sigma^+$-polarized pulses.

Figure 2 shows typical experimental co-circularly polarized tr-PL
data for different laser intensities of $I=1, 40, 70$ and $110$ mW
but the same detuning $\delta=-10$ meV. The maximum of the PL signal
increases in a nonlinear manner with the pulse laser intensity (note
that for the weakest intensity $I=1$ mW, the curve has been
amplified by a factor of 225), while the temporal widths reduce
slightly. New features developing on the decay side of the PL are
clearly observed, with a temporal behavior dependent on the pump
laser intensity. In particular, the emergence of a revival of the
decaying PL signal is markedly evident for high laser intensities.
This new peak becomes clearer for high intensity pumping. For
positive detunings the extra peak on the decay side of the tr-PL is
not present (results not shown). These observed intensity-dependent
features can be fitted with good quantitative agreement using our
theoretical model with memory effects controlled by the pulsed pump
intensity.

\begin{figure}[tbh]
\centerline{ {\includegraphics
[height=10.5cm,width=9.0cm]{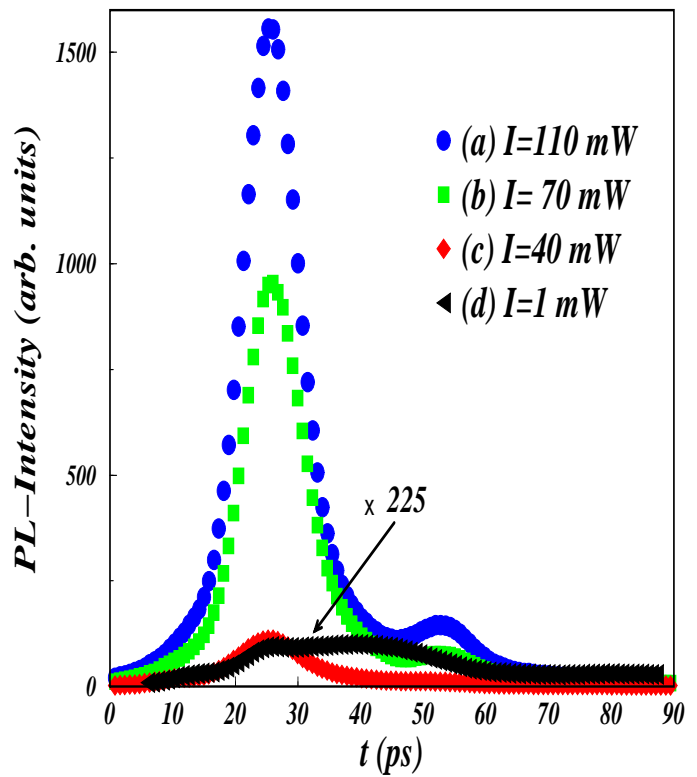}}} \caption{\label{fig-2}
Experimental co-circularly polarized tr-PL signals for a detuning
$\delta=-10$ meV and different pulse laser intensities: I=110 mW
(blue  circles), I=70 mW (green squares),
 I=40 mW (red diamonds) and  I=1 (mW) (black triangles).}
\end{figure}

\subsection{Discussion}

From the theoretical side our aim is to calculate the time evolution
of the mean number of polaritons in the optically active
$\overrightarrow{k}=0$ state, i.e. $<a_0^{\dag}a_0>(t)$, by
numerically solving Eq.(\ref{Eq:ef}). The parameters used correspond
to CdTe microcavities with an exciton mass
$M_X=m_e+m_h=(0.51+0.9)m_0=1.41m_0$. We consider only heavy-hole
excitons coupled to the cavity mode.

The relaxation of UP-resonantly created polaritons can be
qualitatively described as follows. The direct relaxation to the LP
states is strongly inhibited and the majority of polaritons scatter
to large-k LP states. The situation is equivalent to that obtained
after non-resonant excitation, i.e. there is a large polariton
population at the LP bottleneck; from there polaritons relax to
$k\sim0$ states via polariton-polariton parametric scattering. We
assume that the accumulation of intermediate or "bottleneck"
polaritons, following the actual laser pulse, takes place for a
longer period of time (tens of picoseconds) than the exciting pulse
width ($\sim$ 2 ps), corresponding to an asymmetrical shape. Thus,
it is to be modeled by an effective pump-polariton field of the form
$h(t)=At^3e^{-\beta t}$, where $\beta$ can be associated with the
accumulation rate of intermediate polaritons. This assumption is
justified since time-resolved PL in such CdTe microcavities, under
pulsed excitation, has a clear asymmetric shape with typical decay
times on the order of $10^1-10^2$ ps. We will call as the effective
intensity, the square of the time integral of the polariton-pump
pulse, $I_{pp}\sim |A\beta^{-4}|^2$, where dimensionless $I_{pp}$ is
assumed to be directly linked to the intensity of the laser pulse,
$I$, and will enable us to connect the real excitation pump
intensity with the intermediate-polariton-pump classical field. In
all cases reported below, we focus on the analysis of co-polarized
signals, i.e. the excitation is $\sigma^+$-polarized and the
emission is analyzed into its $\sigma^+$-polarized component.

By energy conservation requirements, the parametric-like interaction
(see Eq.(\ref{Eq:e4})), giving rise to the simultaneous creation of
a $\overrightarrow{k}=0$ polariton and a bath polariton, makes only
sense for negative values of $\delta$. Besides that, tr-PL
experimental results show unconventional oscillatory features in
co-polarized signals only for negative values of $\delta$ where a
polariton trap is possible\cite{lola2}. Consequently, we restrict
our discussion and comparison with experimental data to the $\delta
< 0$ case.

Figure 3 shows the high-intensity experimental tr-PL signals from
Fig.2 (symbols, now independently normalized to their maximum value
for each pulse intensity), with the corresponding fits (solid lines)
to our memory-based theory. A non-oscillatory decay is obtained in
the weak pump limit or when memory effects are neglected (see
below). We therefore focus on the high intensity laser pumping
cases. The memory kernel $K_0(\tau)$ depends on the coupling
function $g(\vec k,\vec k')$ through its width (in $k$-space)
$\sigma_{0}$ and strength $\Gamma^{1/2}$. We use the following
fitting parameters:  $\delta=-10$ meV, $\sigma_0=25.7\times 10^7
m^{-1}$, $\Gamma=3\times 10^{24} s^{-2}$,
which together with the bare exciton mass, $M_X$, define a typical
signal-idler coupling time scale as $t_c=e^{|\delta|/\alpha}
\sqrt{|\delta|\alpha/4\pi\Gamma^2}\sim 10^{-10} s$. In terms of this
coupling time, the effective-pump inverse temporal width is taken as
$\beta=6/t_c$. Other parameters are $\Gamma_0=0.45 |\delta|$ and
$V_0=0.02 |\delta|$. By adjusting the pump-polariton pulse amplitude
$A$, which together with $\beta$ determine $h(t)$ (insets in Figure
2), in such a way to obtain the same relative ratios between the
effective pump-polariton intensities, $I_{pp}=1, 0.64, 0.36$ as
those in the actual laser pulses $I=110, 70, 40$ mW, the computed
tr-PL results reproduce both qualitatively and quantitatively the
experimental data. A non-linear initial rise of the tr-PL signal,
instead of a simple linear one, as well as a temporal width that
decreases for increasing pump intensity, are clear signatures of
non-Markovian behavior. Clearly, at long times the decays are
similar for both high- and low-intensity pump levels, thus memory
effects are practically unobservable in that limit. The
non-Markovian effects are more visible after the accumulation of
intermediate polaritons reaches its maximum value.

\begin{figure}[tbh]
\centerline{ {\includegraphics
[height=10.5cm,width=9.0cm]{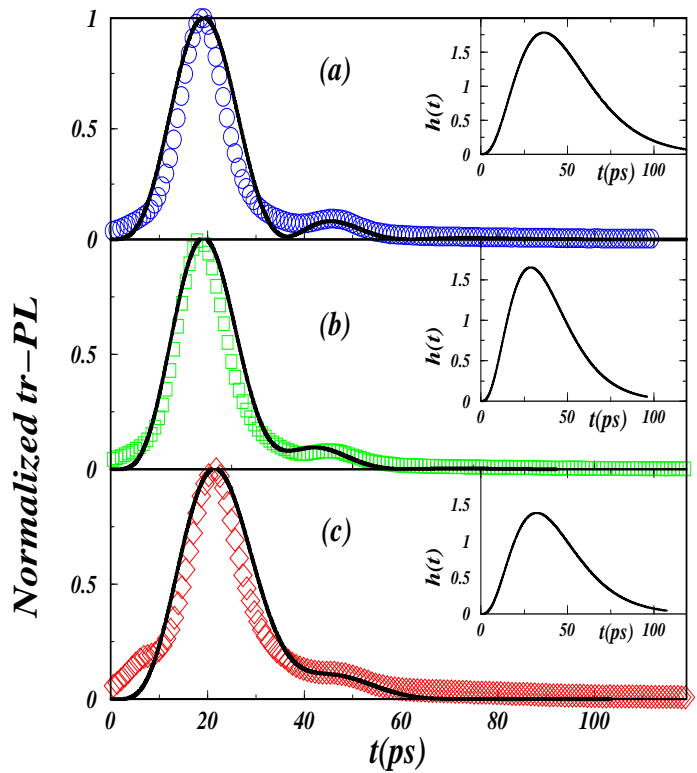}}} \caption{\label{fig-3}
Experimental tr-PL normalized signal for different pump intensities
(symbols) and theoretical fits (continuous lines). Experimental
pulse laser intensities: (a) I=110 mW, (b)=70 mW and (c) I=40 mW.
Insets represent the effective pump-polariton pulses $h(t)$. For
theoretical parameters see the text.}
\end{figure}

In order to explore additional consequences of our model, beyond the
comparison with measured tr-PL data, we proceed to discuss the
effects of the pump-polariton pulse parameters on the tr-PL results.
Figure 4 shows simulations of tr-PL, where only variations of the
polariton-pump intensities $I_{pp}$ are assumed ($I_{pp}=1$
corresponds to the fitting pump intensity for the experimental curve
in Fig.3-a). Coupling parameters $\sigma_{0}$, $\Gamma$ and $V_0$
are identical to those used in Fig. 3. Calculated tr-PL signals are
plotted in Figure 4-a on a logarithmic vertical scale to show the
sensitivity of memory effects on the pump-polariton intensity or
equivalently on the laser pulse intensity. Oscillations in the
polariton population develop as the polariton-pump intensity
increases. Our results clearly demonstrate that the polariton
population dynamics, for a high pump excitation, shows a
non-exponential (oscillatory) behavior in contrast with the typical
\begin{figure}[tbh]
\centerline{ {\includegraphics [height=8.5cm,width=8.8cm]{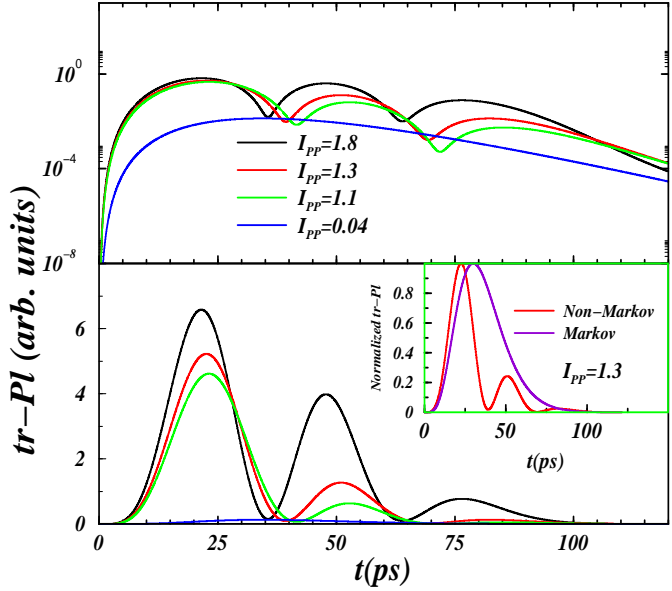}}}
\caption{\label{fig-4} Simulated tr-PL signals for different
pump-polariton intensities $I_{pp}$. (a) Logarithmic scale. (b)
Linear-scale. Inset: Comparison between Markovian (delta kernel) and
non-Markovian tr-PL results for a high intensity pulse $I_{pp}=1.3$
corresponding to a laser intensity of $I=143$ mW.}
\end{figure}
exponential one predicted by simple Markovian approaches. In the
low-intensity limit a Markovian approach should be sufficient to
explain the non-oscillatory behavior. This is further illustrated on
Fig.4-b, where a vertical linear scale is used to better see the
evolution towards a typical Markovian signal for low intensity
pumping. Furthermore, by collapsing the memory function or kernel
(see Eqs.(\ref{Eq:ker}) and (\ref{Eq:kernel})) to a delta function,
i.e. $K(\tau)\sim \delta(\tau)$, for a high pump intensity case, the
peak on the decay side of the tr-PL disappears, as it is
demonstrated in the inset of Figure 4(b). This fact brings further
support to an intensity-controlled memory mechanism in these CdTe
microcavities.

One last issue to address is the sensitivity of our results on the
effective pump-pulse parameters, $A$ corresponding to the pulse
height and $\beta$ associated to the inverse decay time. In Figure 5
simulated tr-PL results, for a given effective pump intensity
$I_{pp}=1$ but different pulse parameters, are depicted. We
emphasize that the coupling strength between signal and idler
subsystems is time-dependent and goes as $\Gamma^{1/2}h(t)$. The
main conclusion from Figure 5 is that non-Markovian signatures are
enhanced for an impulse-like effective pump pulse, i.e. for large
$\beta$ or equivalently, for given $I_{pp}$, a large amplitude $A$.
A non-oscillatory behavior of tr-PL signal occurs for wide
pump-pulses but intense and short pump-pulses give rise to
oscillatory patterns in tr-PL. On this basis we can conclude that
memory effects switch-on when a high intensity pump-pulse has a
temporal width such that $\beta t_c^*\gg 1$, where $t_c^*$ is an
intensity controlled coupling time given by $t_c^*=t_c/A^2$. This
feature is fully consistent with our basic premises for the
existence of memory effects in the sense that a large accumulation
rate of intermediate polaritons would lead to a strong coupling
between signal and idler (bath) polaritons. Moreover, our results
suggest new possibilities for indirectly monitoring the rapid
relaxation of UP polaritons to LP states. Non-Markovian effects in
tr-PL could be interpreted as a signature of a rapid relaxation
dynamics from nonresonantly pumped polaritons to optically active
lower branch polaritons.

Markovian theories give temporal broader tr-PL signals and no
oscillations are observed. Some previous theoretical treatments
\cite{shelykh} have attributed those oscillations to the existence
of other states as dark excitons or spin-splitted states. Our
\begin{figure}[tbh]
\centerline{ {\includegraphics [height=8.5cm,width=8.8cm]{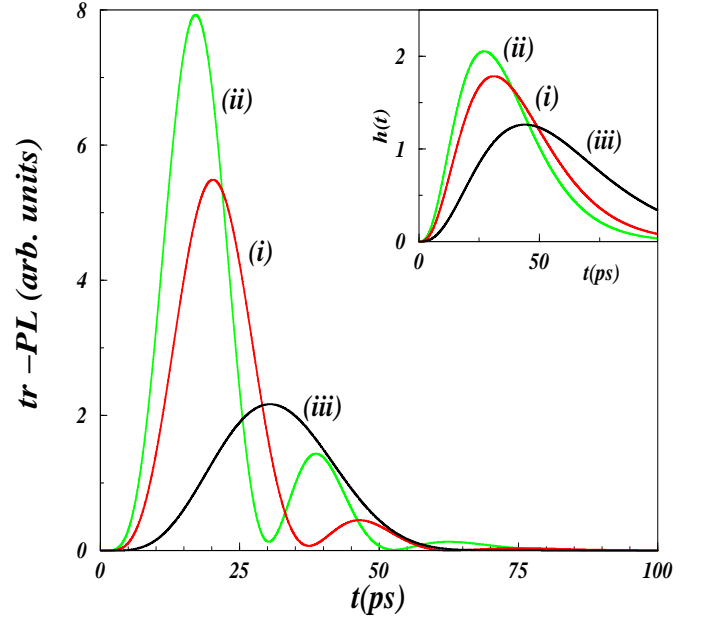}}}
\caption{\label{fig-5} Simulated tr-PL signals for a fixed
polariton-pump effective intensity $I_{pp}=1$, or equivalently
$I=110$ mW, but different pulse-shape parameters: (i) Red curve,
$A=A_1$, $\beta=\beta_1$, corresponding to those parameters used in
the fitting of the experimental data in Fig.3-a. (ii) Green curve,
$A=7A_1/4$, $\beta=(7/4)^{1/4}\beta_1$. (iii) Black curve $A=A_1/4$,
$\beta=\beta_1/\sqrt{2}$. The inset shows the pump-pulses h(t).}
\end{figure}
theoretical analysis shows that oscillations in the tr-PL can be
also produced by non-Markovian effects coming from the coupling
between the trapped optically active polaritons with a polariton
bath via a modulated parametric-like interaction. Furthermore, the
wide range of intensities for which a satisfactory agreement is
obtained demonstrates that our theoretical model indeed captures the
main physics of the pulsed response of polaritons in such II-VI SM
as due to memory effects, without the need of invoking other states
as responsible for those oscillations, as required by Markovian
theories.

Finally, it is worthwhile to mention that spin effects are important
in the polariton physics for such II-VI SM. For the cross-polarized
case, i.e. excitation $\sigma^+$-polarized and emission analyzed
into its $\sigma^-$ component, the same set of parameters fitting a
co-polarized case, for identical detuning and pump intensity, does
not fit the experimentally observed results. This evidences that in
this case the spin dependent polariton scattering must be properly
included to describe the observed data.

\section{Conclusions}

In summary, we have demonstrated that non-Markovian, or memory
effects, produce oscillatory features in tr-PL signals in II-VI SM.
In particular, we have found that the nonlinear rise and the revival
of the decaying PL signal for high laser intensities is explained in
terms of a non-Markovian behavior of the optically active polariton
system as a consequence of being efficiently coupled to a structured
reservoir. These new features are enhanced in the high-laser-power
excitation case. The shape and temporal width of a pump laser pulse
should lead to control the dynamics of relaxing polaritons from a
nonresonant initial distribution in the UP to access optically
active states in the LP branch. Experiments to investigate this
control effect would provide valuable insight into polariton
dynamics and are feasible with current technology.

It is important to note, that while the tr-PL signal does not give a
definite answer to the question about the exact intermediate states
of the relaxation process from the initially created UP polaritons,
the qualitative and quantitative agreement with the tr-PL
experimental results, as a function of the excitation intensity, are
strong indications of the validity of the present model. A full
quantum treatment of the intermediate polariton states would be
desirable. These analysis, which require considering a much more
complex quantum state space, are left for further investigations.

\section{Acknowledgments}
LQ would like to acknowledge MEC(Spain) for a sabbatical grant and
the Universidad Aut\'onoma de Madrid-Spain for hospitality. FJR and
LQ have been partially supported by research project funds from
Facultad de Ciencias-Uniandes. MDM thanks the Ram\'on y Cajal
Program. This work was partially supported by the Spanish MEC
(MAT2005-01388, NAN2004-09109-C04-04, and QOIT-CSD2006-00019), and
the CAM (S-0505/ESP-0200).

\end{document}